# Some exact solutions for localized and periodic structures in active optical cavity

A.Yu.Okulov

*P.N.Lebedev Physical Institute of Russian Academy of Sciences*
*Leninsky prospect 53, 119991 Moscow, Russia , e_mail: okulov@sci.lebedev.ru*

**Abstract.** The nonlocal order parameter equation for nonlinear optical resonator is approximated in the limit of high Fresnel number by the Swift- Hohenberg equation. The exact soliton and periodic solutions of the Swift - Hohenberg equation for optical resonator with thin slice gain element are obtained. It is shown that hyperbolic "flat" secant, "chirped" secant and Jacobi elliptic functions solutions in one transverse dimension are nonlinear eigenmodes for plane - parallel, confocal and quasiconfocal geometries.

1. **Introduction.**

   Transverse modelocking had been considered since late 60-th [1] as effective tool for control of laser emission parameters. In linear regime the localized structures are gaussian wave packets bouncing in transverse section:

$$E(x, y, t) = E_0 \, exp\left[-\left(x - x_0 \cos(\Omega t)\right)^2\right] / \sqrt{\pi} \qquad (1).$$

The nonlinear mechanisms of modelocking[2], considered initially for parametric solitons[3] and transverse optical bistability problems [4] are described inevitably by hyperbolic secant and Jacobi elliptic functions [5] solutions. The alternative approach exploits the concept of so-called "autosolitons" as bounded states of switching waves [6].

   In present paper the approach based on "nonlocal maps"[5,7] is being developed. Using variational principle the nonlocal evolution equation is developed in order to take into account the boundary conditions in explicit form. Under the proper choice of the moments of gain and refractive index distributions the Swift-Hohenberg equation obtained. Using the methods developed by Akhmediev et al. [8], the exact solutions in the form of hyperbolic and elliptic functions are obtained.

2. **Order parameter equation.**

   We use the gradient form of dynamical equation [9]:

$$i\hbar \frac{\partial E}{\partial t} = -\frac{\delta \Phi}{\delta \Psi} \qquad (2)$$

With free energy functional being nonlocal on spatial coordinates:

$$\Phi_{nonloc}(z) = \iint \left[\left\{K(\vec{r} - \vec{r}', z) E_n(\vec{r}') E_n^*(\vec{r}) + K^*(\vec{r} - \vec{r}', z) E_n^*(\vec{r}') E_n(\vec{r})\right\} d^2\vec{r}'\right] d^2\vec{r} +$$
$$+ \int \left\{\frac{\alpha |E|^2}{2} + \frac{\beta |E|^4}{4} + \frac{\gamma |E|^6}{6}\right\} d^2\vec{r} \qquad (3)$$

In discrete time $n$ approximation, with time interval $(2L_R / c)$ we have :

$$E_{n+1}(\vec{r}, z) = -\frac{\delta \Phi_{nonloc}}{\delta E_n} \qquad (4)$$

The time – dependent order parameter equation takes the form of nonlocal map [5, 7]:

$$E_{n+1}(\vec{r}, z) = \iint K(\vec{r} - \vec{r}', z) \, f\left[E_n(\vec{r}')\right] d^2\vec{r}' \qquad (5)$$

Where nonlocal kernel is given by Green function for free - space parabolic wave equation[5]:

$$K(\vec{r} - \vec{r}', z) = \frac{ik}{2\pi z} \exp(ikz) \, R(x,y) \, \exp\left\{\frac{ik(\vec{r} - \vec{r}')^2}{2z}\right\} \quad (6)$$

3. **Model.**

The model uses thin – slice two – level saturable medium (fig. 1) approximation relevant to diode-pumped microchip solid-state lasers and broad area surface emitting lasers [10]:

$$f(E_n) = E_n \frac{\sigma N_0(\vec{r}) L_a (1 - i\Delta\omega T_2)}{2(1 + \sigma T_1 c |E_n|^2 / 8\pi\hbar\omega)} + E_n \quad (7)$$

where $\sigma$ - stimulated emission cross section, $N_0$ - density of resonant ions in dielectric of electron-hole pairs in semiconductor, $\Delta\omega$ - detuning from resonance, $T_2$ - inverse linewidth, $T_1$ - upper level relaxation time, $c$ -speed of light, $L_a$ - thickness of the gain medium ($L_a \ll L_R$). The nonlocality is taken into account by expanding of the field under convolution integral (5) in Taylor series on powers of inverse Fresnel number $N_F^{-1} = L_R / (k D^2)$.

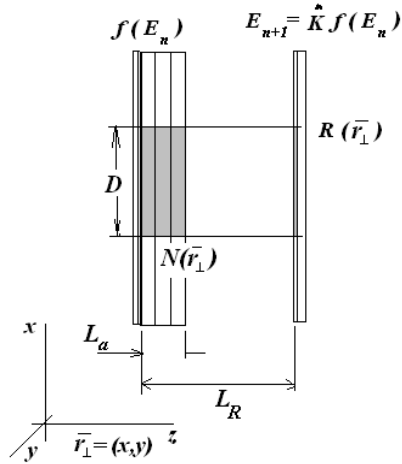

**FIGURE 1.** Geometry of Fabry-Perot thin slice laser.

Spatial filtering on nonlinear diafragm of width $D$ is described by complex diffusion terms in resulting Swift-Hohenberg equation in the limit of large $D$:

$$\frac{\partial E}{\partial t} + M_0 E + M_2 \Delta_\perp E + M_4 \Delta^2_\perp E = E_n \frac{\sigma N_0 L_a c (1 - i\Delta\omega T_2)}{2 L_R (1 + \sigma T_1 c |E_n|^2 / 8\pi\hbar\omega)} \quad (8)$$

The all moments of gain distribution are given exactly. For symmetric diaphragm the first three nonzero moments have the following form:

$$M_0 = \tau_R^{-1}; \quad M_2 = \frac{i L_R}{2 k \tau_R}; \quad M_4 = -\frac{3 L_R^2}{2 k^2 \tau_R}; \quad (9)$$

The following exact localized solution had been obtained by approach developed by Akhmediev et al [8] in the form of chirped hyperbolic secant (fig.2):

$$E(x,t) = \chi \{ \mathrm{sech}(\chi x) \exp(i\theta \mathrm{Log}[\mathrm{sech}(\chi x)]) + \vartheta \} \exp(i\Omega t) \quad (10)$$

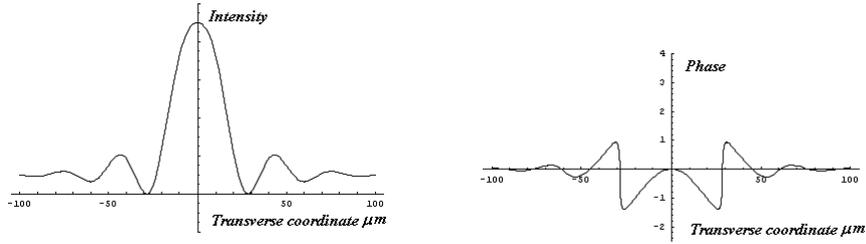

**FIGURE 2.** Distribution of intensity (left) and phase (right) for exact solution (10).

The 2D localized vortex solution had been obtained in similar form using anzatz from /9,13/(fig.3):

$$E(x,y,t) = B + A\, sech(x^2 + y^2 / w)(x^2 + y^2)^{S/2}$$
$$exp(i(log[sech(x^2 + y^2 / w)]))exp(i\,S\,arg[y/x]) + i\Omega t))  \quad (11)$$

$$E(x,y,t) = \frac{A}{(B + ch(\alpha x + \beta y + \Omega t + \zeta))} exp(i(\alpha x + \beta y + \Omega t))  \quad (18)$$

**FIGURE 3.** Exact two-dimensional solution of Ginzburg-Landau equation in 2+1. Distribution of intensity versus transverse coordinates $x$, $y$.

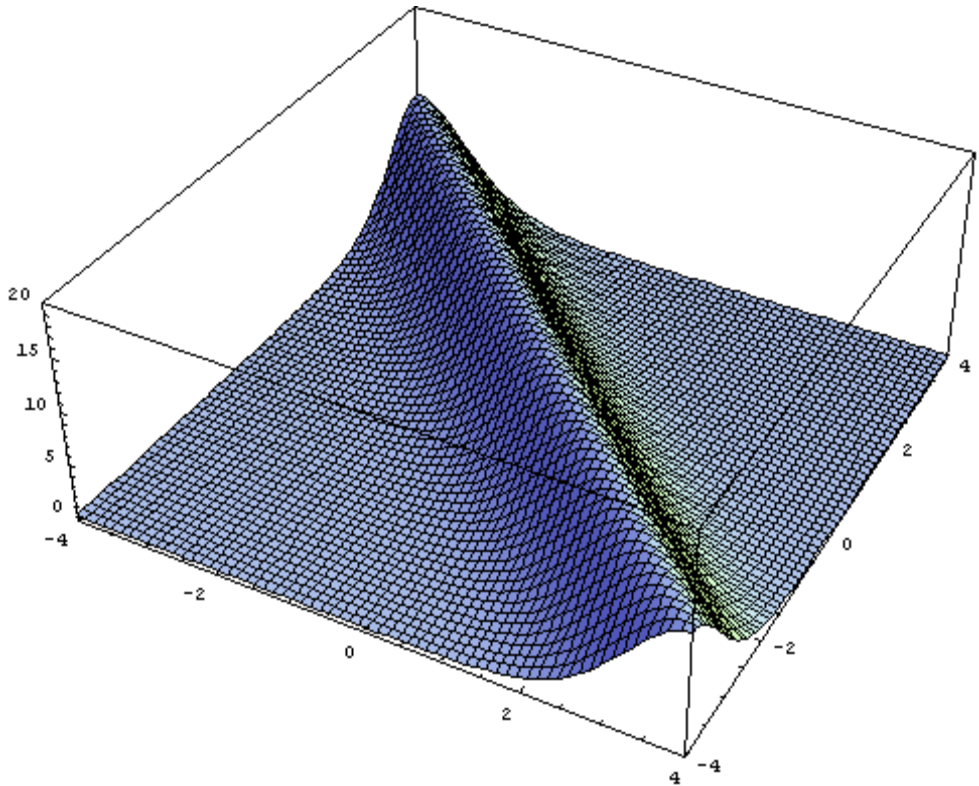

The 2D localized vortex solution had been obtained in the form "chirped" hyperbolic secant /11/(fig.4):

$$E(x,y,t) = B + A\,\mathrm{sech}(x^2+y^2/w)(x^2+y^2)^{S/2}$$
$$\exp(i(\log[\mathrm{sech}(x^2+y^2/w)]))\exp(iS\arg[y/x]) + i\Omega t)) \qquad (11)$$

**FIGURE 4.** Distribution of intensity for 2D vortex solution with unit topological charge *S = 1.*

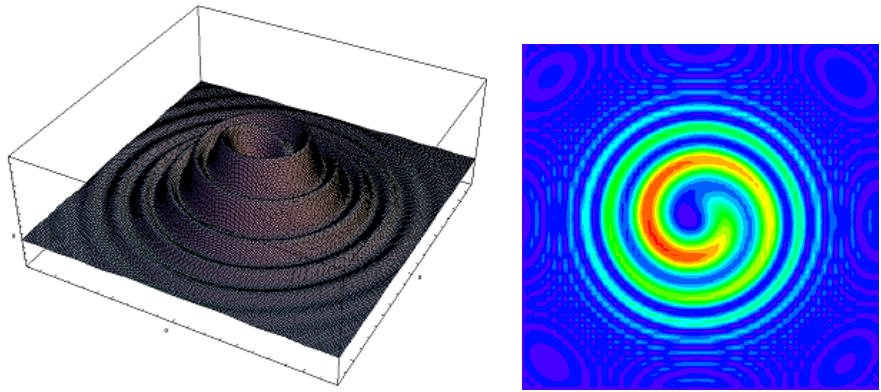